\documentclass[paper,twocolumn,showpacs,superscriptaddress,apl]{revtex4-1}
\usepackage{amsfonts}
\usepackage{amsmath}
\usepackage{amssymb}
\usepackage{graphicx}

\begin{document}
\title{Control of absorption of monolayer MoS$_{2}$ thin-film transistor in one-dimensional defective photonic crystal}

\author{Fang-Fang Yang}
\address{Nanoscale Science and Technology  Laboratory,  Institute for Advanced Study, Nanchang University, Nanchang 330031, China}

\author{Ying-Long Huang}
\address{BOE Technology Group Co., LTD, Dize Road, BDA, Beijing 100176, China}

\author{Wen-bo Xiao}
\address{Key Lab of Non destructive Test (Ministry of Education), Nanchang Hang Kong University, Nanchang 330063, China}

\author{Jiang-Tao Liu}
\email{jtliu@semi.ac.cn}
\address{Nanoscale Science and Technology  Laboratory,  Institute for Advanced Study, Nanchang University, Nanchang 330031, China}
\address{Department of Physics, Nanchang University, Nanchang 330031, China}

\author{Nian-Hua Liu}
\email{nhliu@ncu.edu.cn}
\address{Nanoscale Science and Technology  Laboratory,  Institute for Advanced Study, Nanchang University, Nanchang 330031, China}
\address{Department of Physics, Nanchang University, Nanchang 330031, China}





\date{\today}

\begin{abstract}
The light absorption and transmission of monolayer MoS$_{2}$ in a one-dimensional defective photonic crystal (d-1DPC) is theoretically investigated. The study shows that the strong interference effect decreases photon density in particular areas of the microcavity. The d-1DPC can reduce light absorption of monolayer MoS$_{2}$ and enhance light transmission. The impact of monolayer MoS$_{2}$ light absorption on the localization effect of photon is investigated when monolayer MoS$_{2}$ and the organic light-emitting diode are located in the same microcavity. However, monolayer MoS$_{2}$ does not reduce the localization effect of light by regulating the position of monolayer MoS$_{2}$ in the microcavity.
\end{abstract}


\maketitle

A transparent and flexible thin-film transistor (TFT) is the foundation of next-generation display technology. Traditional transparent and flexible TFT is mostly based on oxide semiconductors,  organic materials, and carbon nanotubes\cite{ RPP09AJ, S13DMS, CSR10HK}. Recently, monolayer MoS$_{2}$ TFT has elicited considerable attention \cite{NN12QHW}. Monolayer MoS$_{2}$ exhibits good electrical and mechanical properties. TFT made of monolayer MoS$_{2}$ consume very low power with high on/off ratio of up to 10$^{5}$\cite{NL12JP, AN13GAS,APL13JP, AN13GHL, S13JY}, which is significantly higher than that of graphene thin-film transistors\cite{NL12SKL}. Given its excellent flexibility, the electrical property of monolayer MoS$_{2}$ film transistors slightly changes even after bending for hundreds of times at 0.75 mm radius \cite{NL12JP}.

However, monolayer MoS$_{2}$ shows higher light absorption ($\sim$10\%). The transmittance of monolayer MoS$_{2}$ is only 70\% to 90\%.
 Monolayer MoS$_{2}$ conductivity is very sensitive to light \cite{NN13OLS}.
 Hence, it is an efficient material for photoelectric detectors.  Notably, the backlight absorption of monolayer MoS$_{2}$ affects its conductivity, thereby decreasing the on/off ratio. These features are unfavorable for creating transparent monolayer MoS$_{2}$ TFT with light irradiation. Transparent TFT can extend the application scope of display technology, increase the opening ratio of display elements, and reduce power consumption.

Therefore, light absorption and transmission should be regulated for creating transparent monolayer MoS$_{2}$ film transistors. Many researchers have investigated the optical regulation of two-dimensional materials, such as graphene and monolayer MoS$_{2}$. Previous studies mostly focused on different optical microstructures to enhance the absorption of two-dimensional materials \cite{PRL12ST, PRB12AF, OL13MAV, AP14JZ,EPL13NMRP, PRB08ZZZ, APL13XZ, APL12JTL,APL13QY}. For example, the one-dimensional defective photonic crystal (d-1DPC) microcavity can be used to significantly enhance two-dimensional materials absorption due to the localization effect of light in microcavity \cite{PRB12AF, OL13MAV, AP14JZ}. This enhancement is mostly based on the interference effect of light. Light interference can usually enhance the absorption of two-dimensional materials in particular areas and reduce their absorption in other areas. Therefore, d-1DPC can be used similarly to reduce the absorption of two-dimensional materials. Flexible photonic crystal products are also produced from industrialization\cite{APL05LH}. As such, composite structures composed of flexible 1DPC and monolayer MoS$_{2}$ are expected to exhibit excellent mechanical properties.

\begin{figure}[b]
\centering
\includegraphics[width=0.95\columnwidth,clip]{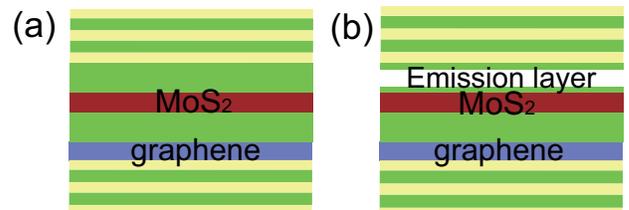}
\caption{(Color online) (a) Schematic of defective photonic crystal-monolayer MoS$_{2}$ film transistor composite structure. (b) Schematic of defective photonic crystal-OLED-monolayer MoS$_{2}$ composite structure.
}
\label{fig1}%
\end{figure}

In this paper, the regulatory function of d-1DPC on the light absorption and transmission of monolayer MoS$_{2}$
is investigated theoretically. In d-1DPC, 20-30 times of monolayer MoS$_{2}$ light absorption can be reduced.
Its transmission can also reach up to 98\% and above. However, in d-1DPC, light absorption and transmission of monolayer MoS$_{2}$ are sensitive to the thickness of defective layer and incident angle. To overcome this weakness, we examined the localization effect of light when monolayer MoS$_{2}$ and microcavity organic light-emitting diodes (OLED) \cite{APL05LH,IJOSTIQE02DD,OE11SH,OE13JRK,APL02FJ} are located in the same microcavity. After regulating the position of monolayer MoS$_{2}$ in the microcavity, monolayer MoS$_{2}$ did not reduce the localization effect of light in this area. Hence, integration of monolayer MoS$_{2}$ and OLED can significantly overcome the above mentioned weakness.

The detailed structure is shown in Fig. 1(a). The d-1DPC is formed by (AB)$^{M_{u}}$ACMCGA(BA)$^{M_{d}}$ structure, where A and B refer to the two different kinds of transparent organic films. Their refractive indexes are 1.7 (polyimides) and 1.5 (polymethyl methacrylate)\cite{JPSPA08CAT}, respectively. C refers to the defective layer with a refractive index of 1.5. M and G refer to monolayer MoS$_{2}$ and grid graphene layer. The thickness of the graphene monolayer MoS$_{2}$ is  $d_{g}=0.34$ nm and $d_{MoS_{2}}=0.65$ nm, respectively. The refraction index of graphene layers  in
the visible range $n_{g}=3.0+\mathcal{C}_{1}\frac{\lambda}{3}i$, where $\mathcal{C}_{1}=5.446$ $\mu m ^{-1}$ \cite{APL09MB}.  The permittivity of monolayer MoS$_{2}$ can be extracted from the experiments through the use of two exciton and band transitions \cite{PRL10KFM,JAP14JTL}. The imaginary permittivity of monolayer MoS$_{2}$ can be given by\cite{JAP14JTL}
\begin{eqnarray}
\varepsilon _{i}\left( \omega \right) &=&\frac{f_{ex}^{A}\Gamma _{A}}{(E_{\omega
}-E_{ex}^{A})^{2}+\Gamma _{A}^{2}}+\frac{f_{ex}^{B}\Gamma _{B}}{%
(E_{\omega }-E_{ex}^{B})^{2}+\Gamma _{B}^{2}} \nonumber \\
&+&\frac{f_{b}e}{\hbar\omega}\Theta (E_{\omega } -E_{g}',\Gamma_{band}),
\end{eqnarray}
where  $\Gamma _{A}=28$ meV ($\Gamma _{B}=42$ meV),  $f_{ex}^{A}=0.32$ meV ($f_{ex}^{B}=0.43$ meV), and $E_{ex}^{A}=1.884$ eV ($E_{ex}^{B}=2.02$ eV) are the linewidth, equivalent oscillator strength, and transition energy of  A (B) excitons, respectively,  $E_{g}'=2.43$ eV is the band gap, $E_{\omega
}$ is photon energy,  $f_{b}=59$ is the  equivalent oscillator strength of interband transition,  $\Theta(x,\Gamma)= \frac{1}{\pi }\int_{-\infty }^{x}\frac{\Gamma }{\Gamma ^{2}+\psi^{2}}d\psi $ is the  step function with a broadening of $\Gamma$, where $\Gamma_{band}=0.398$ eV is the linewidth of the interband transition. The real part of the permittivity of monolayer MoS$_{2}$ can be obtained using Kramers-Kronig relations
$\varepsilon _{r}\left( \omega \right) =4.2+\frac{1}{\pi }%
\mathfrak{p}\int_{0}^{\infty }\frac{s\varepsilon_{i}\left( \omega \right) }{%
s^{2}-\omega ^{2}}ds,$
where $\mathfrak{p}$ is the principal value integral.

The standard transfer-matrix method was used for the calculation  \cite{APL12JTL,JAP14JTL}.
In the \emph{l}th layer, the electric field of the TE mode light with incident angle $\theta_{i}$ is given by
\begin{equation}\mathbf{E}_{l}(z,y)=\left[  A_{l}e^{ik_{lz}%
\left(  z-z_{l}\right)  }+B_{l}e^{-ik_{lz}\left(  z-z_{l}\right)  }\right]
e^{ik_{ly}y}\mathbf{e}_{x}, \label{TMM:a1}\end{equation}
and the magnetic field of the TM mode  is given by
\begin{equation}\mathbf{H}_{l}(z,y)=\left[  A_{l}e^{ik_{lz}%
\left(  z-z_{l}\right)  }+B_{l}e^{-ik_{lz}\left(  z-z_{l}\right)  }\right]
e^{ik_{ly}y}\mathbf{e}_{x},\end{equation}
where $k_{l}=k_{lr}+ik_{li}$ is the wave vector of the light, $\mathbf{e}_{x}$ is the unit vectors in the x direction, and $z_{l}$ is the position of the \emph{l}th layer in the z direction.

The electric fields of TE mode or the magnetic fields of TM mode in the (\emph{l}+1)th layer are related to the incident fields by the transfer matrix utilizing the boundary condition \cite{APL12JTL,JAP14JTL}.
Thus, we can obtain  the absorbance of \emph{l}th layer $\mathcal{A}_{l}$  using the Poynting vector $\textbf{S}=\textbf{E}\times\textbf{H}$ \cite{APL12JTL,JAP14JTL}
\begin{equation}
\mathcal{A}_{l}=[{S}_{(l-1)i}+{S}_{(l+1)i}-{S}_{(l-1)o}-{S}_{(l+1)o}]/{S}_{0i},
\end{equation}
where  ${S}_{(l-1)i}$ and ${S}_{(l-1)o}$ [${S}_{(l+1)i}$ and ${S}_{(l+1)o}$] are the incident  and  outgoing Poynting vectors (\emph{l}-1)th [(\emph{l}+1)th] layer, respectively, ${S}_{0i}$ is the incident Poynting vectors in air.

\begin{figure}[t]
\centering
\includegraphics[width=0.95\columnwidth,clip]{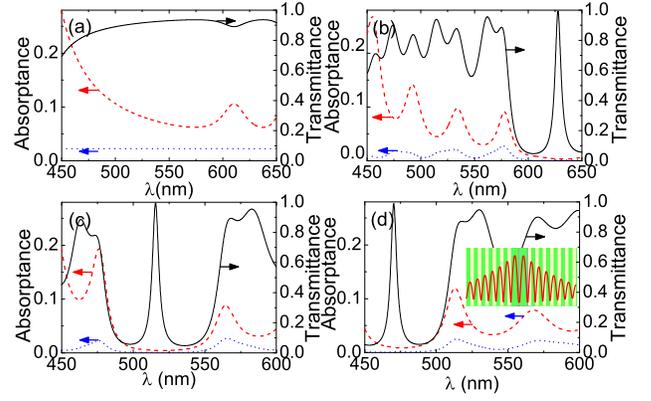}
\caption{(Color online) (a)Transmission (black solid line), absorption (red dashed line) of monolayer MoS$_{2}$, and absorption (blue dots) of graphene in different structures, namely, (a)suspended monolayer MoS$_{2}$,  (b) red d-1DPC, (c) green d-1DPC, and (d) blue d-1DPC. The inset shows the optical field distribution in these structures.
}
\label{fig2}%
\end{figure}

For comparison, we calculated the transmittance and absorption of monolayer MoS$_{2}$, as well as the absorption
of single-layer graphene [Fig. 2(a)]. The absorption of monolayer MoS$_{2}$ is about 6\% to 30\%, and
its transmittance is about 70\% to 90\%. The absorption of graphene is around 2.3\%.
The three color principle was used in general display technology.
The same pixel was decomposed into three colored dots, red, green,
and blue, through a filter or light resource with different colors. We investigated each of the three colors.
The wavelengths of red, blue, and green light are 627, 516, and 470 nm, respectively \cite{APL05LH,IJOSTIQE02DD,OE11SH,OE13JRK,APL02FJ}. 
All the color dots used different photonic crystal-structure parameters with $M_{u}=M_{d}=8$.
The thicknesses of layers A, B, and C for the red photonic crystal are 89.4, 107.3,
and 212.0 nm, respectively. The thicknesses of layers A, B, and C for green photonic
crystal are 73.9, 88.7, and 171.0 nm, respectively. The thicknesses of layers A, B,
and C for blue photonic crystal are 67.1, 80.5, and 158.0 nm, respectively.
The calculation results are shown in Figs. 2a-2d. The corresponding transmittances of red,
blue, and green light can also reach 99.5\%, 99.4\%, and 98.9\%, respectively.
The corresponding absorptions of monolayer MoS$_{2}$ (graphene) in the microcavity are 0.35\% (0.13\%), 0.45\% (0.10\%),
and 0.96\% (0.13\%), which are 20.2 times (17.7 times), 19.8 times (23 times), and 17.8 times (17.7 times) less than those in the non-microcavity monolayer MoS$_{2}$ (graphene). A strong localization effect of light occurred in the defective area of photonic crystal [inset of Fig. 2(d)]. However, the light intensity is largely weakened in the special defective position because of the interference effect, thereby causing weak absorption. The microcavity resonance also resulted in a transmittance close to 1. These characteristics can enhance the transmittance of monolayer MoS$_{2}$ film transistors, thereby improving the open ratio of display elements and reducing the effect of light absorption on the conductivity of monolayer MoS$_{2}$ and on/off ratio of film transistors.

\begin{figure}[t]
\centering
\includegraphics[width=0.85\columnwidth,clip]{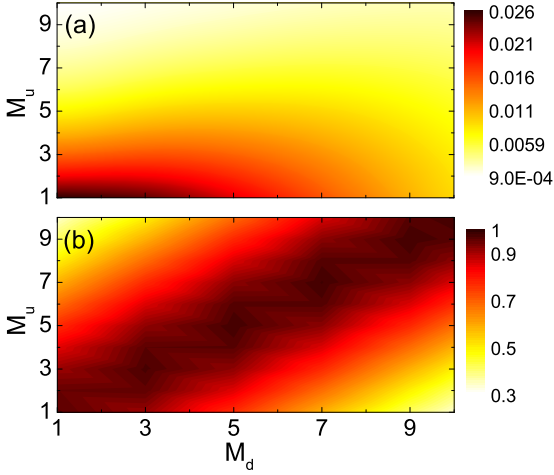}
\caption{(Color online) Effects of the number of cycles ($M_{d}$ and $M_{u}$) in A and B layers on (a) monolayer MoS$_{2}$ absorption and (b) d-1DPC transmission.
}
\label{fig3}%
\end{figure}

Changing the number of cycles ($M_{u}$ and $M_{d}$) of A and B layers can regulate the absorption of monolayer MoS$_{2}$ (Fig. 3).
For example, in green light, when the numbers of cycles in layers A and B increase simultaneously,
the interference effect is more significant, but the absorption is lower. For $M_{u}=M_{d}=7$,
the absorption of monolayer MoS$_{2}$ is about 0.58\%; for $M_{u}=M_{d}=10$,
the absorption of monolayer MoS$_{2}$ is only 0.28\%, which is reduced by 31.8 times. In an asymmetric case ($M_{u}\neq M_{d}$), the absorption and transmittance of monolayer MoS$_{2}$ (graphene) are both reduced.

Resonance in the defective microcavity of photonic crystal is sensitive to the thickness of the defective layer and incident angle (Fig. 4). The resonance wavelength of the microcavity meets  $M_{0}\lambda_{c}=L_{c}\cos\theta'$;   $L_{c}=n_{c}2d_{c}$ refers to the optical path length of the microcavity, $n_{c}$ and $d_{c}$ is the refractive index and thickness of C layer; $M_{0}$ is a positive integer; and  $\theta'=\arcsin\theta_{i}$ refers to the propagation angle in the defective layer, $\theta_{i}$ is the incident angle. Therefore, when the defective layer becomes thicker, the resonance wavelength will increase linearly. If  $M_{0}$ is an odd number, the absorption of monolayer MoS$_{2}$ will be enhanced, and the maximum absorption can reach 0.44. The corresponding resonance transmittance decreases. If  $M_{0}$ is an even number, the absorption of monolayer MoS$_{2}$ will be inhibited, and the resonance transmittance can reach the maximum. When the incident angle becomes larger, the resonance wavelength shifts to a shortwave. Resonance wavelength is determined by the propagation angle in the defective layer. Therefore, the sensitivity to the incident angle will depend on the index of refraction of the defective layer. If the index of refraction is higher, the resonance peak will change more slightly. The index of refraction of the defective layer is 1.5 (Fig. 4). Thus, the defective layer is sensitive to the incident angle.

\begin{figure}[t]
\centering
\includegraphics[width=0.98\columnwidth,clip]{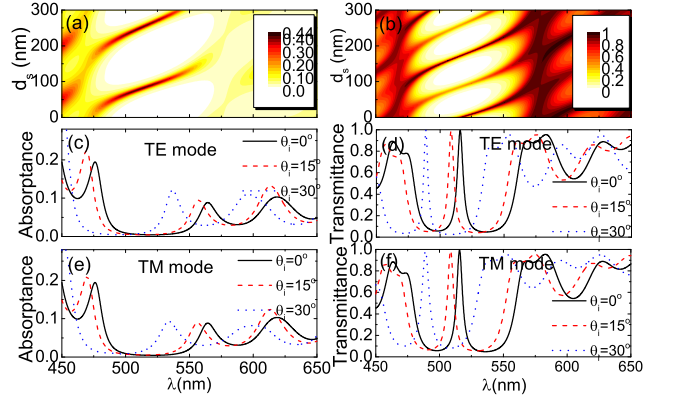}
\caption{(Color online) Influence of the thickness of defective layer on (a) monolayer MoS$_{2}$ absorption and (b) d-1DPC transmission. Effect of incident angle on (c, e) monolayer MoS$_{2}$ absorption and (d, f) d-1DPC transmission.
}
\label{fig4}%
\end{figure}

Resonance in the defective microcavity of photonic crystal is more sensitive to the thickness of the defective layer and the incident angle. The thickness of the defective layer and the incident angle will be changed if the TFT is bent. This flexibility is unfavorable to traditional display technology. However, microcavity-based OLED technology has been widely developing. Considering the localization effect of light, the quantum efficiency of OLED can be significantly improved by integrating OLED in the microcavity, thereby improving brightness and reducing power consumption \cite{APL05LH,IJOSTIQE02DD,OE11SH,OE13JRK,APL02FJ}. If monolayer MoS$_{2}$ thin-film transistor is integrated with OLED in a microcavity, lights emitted by OLED will change synchronously with the change in the thickness of the microcavity and propagation angle of the light, thereby keeping high transmittance and low absorption of monolayer MoS$_{2}$ unchanged. Such integration is also favorable to reduce the monitor thickness. Given the simplified process and reduction of pixel size, the integration can significantly improve the opening ratio of film transistors, which cause low power consumption. 

\begin{figure}[t]
\centering
\includegraphics[width=0.90\columnwidth,clip]{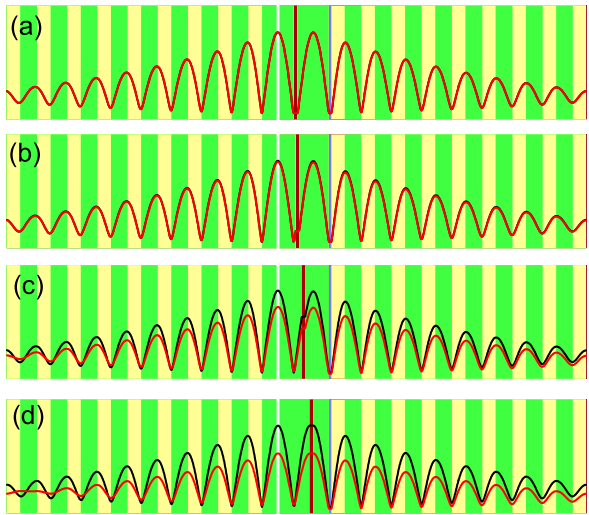}
\caption{(Color online) When OLED and monolayer MoS$_{2}$ are located in the same microcavity, the optical field distribution in d-1DPC as the imaginary part of the monolayer MoS$_{2}$ dielectric constant is either 0 (black solid line) or not 0 (black solid line); (a) monolayer MoS$_{2}$ at the center of microcavity;  monolayer MoS$_{2}$ at (b) 10 nm, (b) 45 nm, and (c) 88.5 nm  away from the center.
}
\label{fig5}%
\end{figure}

Therefore, we investigated the effect of the absorption of monolayer MoS$_{2}$ in the microcavity on light localization. The imaginary part of the monolayer MoS$_{2}$ dielectric constant is either 0 (non-absorption) or not 0 (absorption); the calculation results of optical field distribution are shown in Fig. 5. When monolayer MoS$_{2}$ is located in the minimum value of optical field in the microcavity, light localization is almost unchanged with or without absorption, and the luminescent layer is always in the place with the greatest light intensity. When monolayer MoS$_{2}$ is 10 nm away from the center, the localization effect of the light is almost unchanged. When monolayer MoS$_{2}$ is 45 or 88.5 nm ($\lambda_{c}/4$) away from the center, the absorption becomes strong with significantly reduced localization effect of the light.

Finally, we discuss the achievement of the relative experiment. Organic flexible photonic crystals have been commercially produced before\cite{APL05LH}. The growing and manufacturing technology for monolayer MoS$_{2}$ thin-film transistors also tends to mature\cite{NL12JP, AN13GAS,APL13JP, AN13GHL, S13JY}; particularly, the method developed by G. A. Salvatore el at. can effectively transfer monolayer MoS$_{2}$ TFT to integrate with other microstructures \cite{AN13GAS}. Microcavity OLED-related manufacturing technologies, including microcavity integration, manufacturing, and connecting of electrodes, also continue to improve \cite{APL05LH,IJOSTIQE02DD,OE11SH,OE13JRK,APL02FJ}; particularly, three color OLED with organic flexible d-1DPC have been produced \cite{APL05LH}.  Hence, the development of microcavity-integrated monolayer MoS$_{2}$ film transistor and display technology of OLED is completely feasible.

In conclusion, the regulatory function of d-1DPC in the absorption and transmission of monolayer MoS$_{2}$ is investigated.
 Given the strong interference effect, photon density in particular areas of the microcavity decreased.
 In d-1DPC, the absorption of monolayer MoS$_{2}$ can be reduced by more than 30 times, and transmittance can be more than 98\%.
  When monolayer MoS$_{2}$ and OLEDs are located in the same microcavity, the absorption of monolayer MoS$_{2}$ is low and does not reduce the localization effect of the light in the microcavity by regulating the position of monolayer MoS$_{2}$ in the microcavity. This study will have important applications in the field of monolayer MoS$_{2}$ and OLED-integrated monitors.

This work was
supported by the NSFC (Grant Nos. 11364033 and 11264030).


\begin{thebibliography}{10}

\bibitem{RPP09AJ}
Anderson Janotti and Chris G~Van de~Walle.
\newblock Fundamentals of zinc oxide as a semiconductor.
\newblock {\em Rep. Prog. Phys.}, 72:126501, 2009.

\bibitem{S13DMS}
Wen-Cai Ren Hui-Ming~Cheng Dong-Ming~Sun, Chang~Liu.
\newblock A review of carbon nanotube- and graphene-based flexible thin-film
  transistors.
\newblock {\em Small}, 9:1188--205, 2013.

\bibitem{CSR10HK}
Hagen Klaukas.
\newblock Organic thin-film transistor.
\newblock {\em Chem. Soc. Rev.}, 39:2643--2666, 2010.

\bibitem{NN12QHW}
Q.~H. Wang, K.~Kalantar-Zadeh, A.~Kis, J.~N. Coleman, and M.~S. Strano.
\newblock Electronics and optoelectronics of two-dimensional transition metal
  dichalcogenides.
\newblock {\em Nat. Nanotech.}, 7:699--712, 2012.

\bibitem{NL12JP}
J.~Pu, Y.~Yomogida, K.~K. Liu, L.~J. Li, Y.~Iwasa, and T.~Takenobu.
\newblock Highly flexible {MoS$_{2}$} thin-film transistors with ion gel
  dielectrics.
\newblock {\em Nano Lett.}, 12:4013--4017, 2012.

\bibitem{AN13GAS}
Giovanni~A. Salvatore, Niko M\"{u}nzenrieder, Cl\'{e}ment Barraud, Luisa Petti,
  Christoph Zysset, Lars B\"{u}the, Klaus Ensslin, and Gerhard Tr\"{o}ster.
\newblock Fabrication and transfer of flexible few-layers mos2 thin film
  transistors to any arbitrary substrate.
\newblock {\em ACS Nano}, 7:8809--8815, 2013.

\bibitem{APL13JP}
Jiang Pu, Yijin Zhang, Yoshifumi Wada, Jacob Tse-Wei Wang, Lain-Jong Li,
  Yoshihiro Iwasa, and Taishi Takenobu.
\newblock Fabrication of stretchable mos2 thin-film transistors using elastic
  ion-gel gate dielectrics.
\newblock {\em Appl. Phys. Lett.}, 103:023505, 2013.

\bibitem{AN13GHL}
Gwan-Hyoung Lee, Young-Jun Yu, Xu~Cui, Nicholas Petrone, Chul-Ho Lee, Min~Sup
  Choi, Dae-Yeong Lee, Changgu Lee, Won~Jong Yoo, Kenji Watanabe, Takashi
  Taniguchi, Colin Nuckolls, Philip Kim, and James Hone.
\newblock Flexible and transparent mos2 field-effect transistors on hexagonal
  boron nitride-graphene heterostructures.
\newblock {\em ACS Nano}, 7:7931--7936, 2013.

\bibitem{S13JY}
Jongwon Yoon, Woojin Park, Ga-Yeong Bae, Yonghun Kim, Hun~Soo Jang, Yujun Hyun,
  Sung~Kwan Lim, Yung~Ho Kahng, Woong-Ki Hong, Byoung~Hun Lee, and Heung~Cho
  Ko.
\newblock Highly flexible and transparent multilayer mos2 transistors with
  graphene electrodes.
\newblock {\em Small}, 9:3295--3300, 2013.

\bibitem{NL12SKL}
Seoung-Ki Lee, Ho~Young Jang, Sukjae Jang, Euiyoung Choi, Byung~Hee Hong,
  Jaichan Lee, Sungho Park, and Jong-Hyun Ahn.
\newblock All graphene-based thin film transistors on flexible plastic
  substrates.
\newblock {\em Nano Lett.}, 12:3472--3476, 2012.

\bibitem{NN13OLS}
O.~Lopez-Sanchez, D.~Lembke, M.~Kayci, A.~Radenovic, and A.~Kis.
\newblock Ultrasensitive photodetectors based on monolayer {MoS$_{2}$}.
\newblock {\em Nat. Nanotech.}, 8:497--501, 2013.

\bibitem{PRL12ST}
S.~Thongrattanasiri, F.~H.~L. Koppens, and F.~J.~G. de~Abajo.
\newblock Complete optical absorption in periodically patterned graphene.
\newblock {\em Phys. Rev. Lett.}, 108:047401, 2012.

\bibitem{PRB12AF}
A.~Ferreira, N.~M.~R. Peres, R.~M. Ribeiro, and T.~Stauber.
\newblock Graphene-based photodetector with two cavities.
\newblock {\em Phys. Rev. B}, 85:115438, 2012.

\bibitem{OL13MAV}
M.~A. Vincenti, D.~de~Ceglia, M.~Grande, A.~D'Orazio, and M.~Scalora.
\newblock Nonlinear control of absorption in one-dimensional photonic crystal
  with graphene-based defect.
\newblock {\em Opt. Lett.}, 38:3550--3553, 2013.

\bibitem{AP14JZ}
Jiabao Zheng, Robert~A. Barton, and Dirk Englund.
\newblock Broadband coherent absorption in chirped-planar-dielectric cavities
  for 2d-material-based photovoltaics and photodetectors.
\newblock {\em ACS Photonics}, 1:768--774, 2014.

\bibitem{EPL13NMRP}
N.~M.~R. Peres and Yu.~V. Bludov.
\newblock Enhancing the absorption of graphene in the terahertz range.
\newblock {\em EPL}, 101:58002, 2013.

\bibitem{PRB08ZZZ}
Z.~Z. Zhang, Kai Chang, and F.~M. Peeters.
\newblock Tuning of energy levels and optical properties of graphene quantum
  dots.
\newblock {\em Phys. Rev. B}, 77:235411, 2008.

\bibitem{APL13XZ}
Xiaolong Zhu, Wei Yan, Peter~Uhd Jepsen, Ole Hansen, N.~Asger Mortensen, and
  Sanshui Xiao.
\newblock Experimental observation of plasmons in a graphene monolayer resting
  on a two-dimensional subwavelength silicon grating.
\newblock {\em Appl. Phys. Lett.}, 102:131101, 2013.

\bibitem{APL12JTL}
J.~T. Liu, N.~H. Liu, J.~Li, X.~J. Li, and J.~H. Huang.
\newblock Enhanced absorption of graphene with one-dimensional photonic
  crystal.
\newblock {\em Appl. Phys. Lett}, 101:052104, 2012.

\bibitem{APL13QY}
Q.~Ye, J.~Wang, Z.~Liu, Z.~C. Deng, X.~T. Kong, F.~Xing, X.~D. Chen, W.~Y.
  Zhou, C.~P. Zhang, and J.~G. Tian.
\newblock Polarization-dependent optical absorption of graphene under total
  internal reflection.
\newblock {\em Appl. Phys. Lett}, 102:021912, 2013.

\bibitem{APL05LH}
Lintao Hou, Qiong Hou, Yueqi Mo, Junbiao Peng, and Yong Cao.
\newblock All-organic flexible polymer microcavity light-emitting diodes using
  3m reflective multilayer polymer mirrors.
\newblock {\em Appl. Phys. Lett.}, 87:243504, 2005.

\bibitem{IJOSTIQE02DD}
D.~Delbeke.
\newblock High-efficiency semiconductor resonant-cavity light-emitting diodes:
  a review.
\newblock {\em IEEE Journal Selected Topics in Quantum Electronics},
  8:189--206, 2002.

\bibitem{OE11SH}
Simone Hofmann, Michael Thomschke, Bj\"{o}rn L\"{u}ssem, and Karl Leo.
\newblock Top-emitting organic light-emitting diodes.
\newblock {\em Opt. Express}, 19:A1250, 2011.

\bibitem{OE13JRK}
Ja-Ryong Koo, Seok~Jae Lee, Ho~Won Lee, Dong~Hyung Lee, Hyung~Jin Yang,
  Woo~Young Kim, and Young~Kwan Kim.
\newblock Flexible bottom-emitting white organic lightemitting diodes with
  semitransparent ni/ag/ni anode.
\newblock {\em Opt. Express}, 21:11086, 2013.

\bibitem{APL02FJ}
Fr\'{e}d\'{e}rique Jean, Jean-Yves Mulot, Bernard Geffroy, and Christine
  Denisand~Pierre Cambon.
\newblock Microcavity organic light-emitting diodes on silicon.
\newblock {\em Appl. Phys. Lett.}, 81:1717, 2002.

\bibitem{JPSPA08CAT}
Claudio~A. Terraza, Jin-Gang Liu, Yasuhiro Nakamura, Yuji Shibasaki, Shinji
  Ando, and Mitsuru Ueda.
\newblock Synthesis and properties of highly refractive polyimides derived from
  fluorene-bridged sulfur-containing dianhydrides and diamines.
\newblock {\em Journal of Polymer Science Part A: Polymer Chemistry},
  45:1510--1520, 2008.

\bibitem{APL09MB}
M.~Bruna and S.~Borini.
\newblock Optical constants of graphene layers in the visible range.
\newblock {\em Appl. Phys. Lett.}, 94:031901, 2009.

\bibitem{PRL10KFM}
K.~F. Mak, C.~Lee, J.~Hone, J.~Shan, and T.~F. Heinz.
\newblock Atomically thin {MoS$_{2}$}: a new direct-gap semiconductor.
\newblock {\em Phys. Rev. Lett.}, 105:136805, 2010.

\bibitem{JAP14JTL}
Jiang-Tao Liu, Tong-Biao Wang, Xiao-Jing Li, and Nian-Hua Liu.
\newblock Enhanced absorption of monolayer mos2 with resonant back reflector.
\newblock {\em J. Appl. Phys.}, 115:193511, 2014.

\end{thebibliography}
\end{document}